\documentstyle[12pt,aaspp4,tighten]{article}
\def \ref {\noindent\hangindent=1.0in\hangafter=1}
\def \cl {\centerline}

\def\ltsima{$\; \buildrel < \over \sim \;$}
\def\simlt{\lower.5ex\hbox{\ltsima}} 
\def\gtsima{$\; \buildrel > \over \sim \;$}
\def\simgt{\lower.5ex\hbox{\gtsima}} 

\begin{document}

\title{BeppoSAX Observations of Unprecedented Synchrotron Activity
in the BL Lac Object Mkn~501}
 
\author{Elena Pian\altaffilmark{1,2},
Giuseppe Vacanti\altaffilmark{3},
Gianpiero Tagliaferri\altaffilmark{4},
Gabriele Ghisellini\altaffilmark{4},}
 
\author{Laura Maraschi\altaffilmark{5},
Aldo Treves\altaffilmark{6},
C. Megan Urry\altaffilmark{1},
Fabrizio Fiore\altaffilmark{7},
Paolo Giommi\altaffilmark{7},}
 
\author{Eliana Palazzi\altaffilmark{2},
Lucio Chiappetti\altaffilmark{8},
Rita M. Sambruna\altaffilmark{9}}

\altaffiltext{1}{Space Telescope Science Institute, 3700 San Martin Drive, 
Baltimore, MD 21218}
\altaffiltext{2}{ITESRE-CNR, Via Gobetti 101, I-40129 Bologna, Italy}
\altaffiltext{3}{ESA ESTEC, Space Science Department, Astrophysics Division, 
Postbus 299, NL-2200 AG Noordwijk, The Netherlands}
\altaffiltext{4}{Osservatorio Astronomico di Brera, Via Bianchi 46, I-22055
Merate (Lecco), Italy}
\altaffiltext{5}{Osservatorio Astronomico di Brera, Via Brera 28, I-20121
Milan, Italy}
\altaffiltext{6}{Department of Physics, University of Milan at Como, Via 
Lucini 3, I-22100 Como, Italy}
\altaffiltext{7}{SAX SDC, Via Corcolle 19, I-00131, Rome, Italy}
\altaffiltext{8}{IFCTR-CNR, Via Bassini 15, I-20133 Milan, Italy}
\altaffiltext{9}{NASA/Goddard Space Flight Center, Greenbelt, MD 20771}
\altaffilmark{}{electronic addresses: pian@stsci.edu; 
gvacanti@astro.estec.esa.nl;
gtagliaf@astmim.mi.astro.it;
gabriele@merate.mi.astro.it;
maraschi@brera.mi.astro.it;
treves@astmiu.mi.astro.it;
cmu@stsci.edu;
fiore@sax.sdc.asi.it;
giommi@sax.sdc.asi.it;
eliana@tesre.bo.cnr.it;
lucio@ifctr.mi.cnr.it;
rms@latte.astro.psu.edu
}

\begin{abstract}

The BL Lac object Mkn~501, one of the only three extragalactic sources 
(with Mkn~421 and 1ES 2344+514) so far detected at TeV energies, was 
observed with the BeppoSAX satellite on 7, 11, and 16 April 1997 during
a phase of high activity at TeV energies, as monitored with the Whipple, 
HEGRA and CAT Cherenkov telescopes.  Over the whole 0.1-200 keV range 
the spectrum was exceptionally hard 
($\alpha \leq 1$, with $F_\nu \propto \nu^{-\alpha}$) indicating  that 
the X-ray power output peaked at (or above) $\sim$100 keV.  This represents
a shift of at least two orders of magnitude with respect to previous
observations of Mkn~501, a behavior never seen before in this or any other
blazar.  The overall X-ray spectrum hardens with increasing intensity and,
at each epoch, it is softer at larger energies.  The correlated variability 
from soft X-rays to the TeV band points to models in which the same 
population of relativistic electrons produces the X-ray continuum via 
synchrotron radiation and the TeV emission by inverse Compton scattering 
of the synchrotron photons or other seed photons.  For the first time in 
any blazar the synchrotron power is observed to peak at hard X-ray 
energies.  The large shift of the synchrotron peak frequency with respect 
to previous observations of Mkn~501 implies that intrinsic changes in the 
relativistic electron spectrum caused the increase in emitted power.  Due 
to the very high electron energies, the inverse Compton process is limited 
by the Klein-Nishina regime.  This implies a quasi-linear (as opposed to 
quadratic) relation of the variability amplitude in the TeV and hard X-ray 
ranges (for the SSC model) and an increase of the inverse Compton peak 
frequency smaller than that of the synchrotron peak frequency.

\end{abstract}

 
\section{Introduction}

Mkn~501 is one of the closest ($z$=0.034) BL Lacertae objects, and one of 
the brightest at all wavelengths.  It was the second source, after Mkn~421, 
to be detected at TeV energies by the Whipple and HEGRA observatories 
(Quinn et al. 1996; Bradbury et al. 1997).  Historically, its spectral 
energy distribution ($\nu F_\nu$) resembles that of BL Lac objects selected 
at X-ray energies, having a peak in the EUV-soft X-ray energy band.  Such 
objects were defined as High-frequency-peaked BL Lacs, or HBL, by Padovani 
\& Giommi (1995).  In fact, the 2--10~keV spectra observed so far were 
relatively steep, with energy spectral indices $\alpha$ larger than unity 
($F_\nu \propto \nu^{-\alpha}$), meaning the power output peaks below this 
band.  From the EXOSAT data base the hardest X-ray spectrum of Mkn~501, 
measured in one of its brightest states, had a spectral index of $1.2 \pm 0.1$
(Sambruna et al. 1994).  {\it Einstein} measured in two occasions spectral 
indices consistent with values smaller than 1 within the large errors (Urry, 
Mushotzky, \& Holt 1986).  

Mkn~501 was observed with BeppoSAX over a period of $\sim$10 hours on each 7, 
11, and 16 April 1997, during a multiwavelength campaign involving 
ground-based TeV Cherenkov telescopes (Whipple, HEGRA and CAT), plus other 
satellites, CGRO (EGRET and OSSE), RXTE, ISO, and optical telescopes.  First 
results on the TeV observations are presented in Catanese et al. (1997), 
Aharonian et al. (1997a), Barrau et al. (1997).  Infrared, optical, and radio 
data from the multiwavelength campaign, as well as a complete analysis of the 
BeppoSAX data, including a detailed study of the intraday X-ray variability, 
will be presented in forthcoming papers.  Here we concentrate  on the average
spectra obtained during the three BeppoSAX pointings.  In the  X-ray band, 
unprecedented spectral behavior is observed, offering new constraints on 
blazar emission mechanisms.	
 	
\section{Observations, analysis, and results}

A complete description of the BeppoSAX mission is given by Boella et al.  
(1997).  Mkn~501 was observed with the LECS (0.1--10 keV), the MECS 
(1.5--11 keV), and the PDS (13-300 keV).  Event files of the three 
BeppoSAX pointings for the LECS and MECS experiments were linearized 
and cleaned with SAXDAS at the BeppoSAX Science Data Center (SDC; Giommi 
\& Fiore 1997).  Light curves and spectra were accumulated for each 
pointing with the SAXSELECT tool, using 8.5 and 4 arcmin extraction radii 
for the LECS and the MECS, respectively, that provide more than 90\% of 
the fluxes.  The background intensity was evaluated from files 
accumulated from blank fields available at the SDC.  For each of the four 
PDS units, source+background and  background spectra 
were accumulated using the XAS software package, after selecting the source 
visibility windows.  The target was significantly detected up to the highest
energy channels.  Each net spectrum was binned in energy intervals to reach 
a signal-to-noise ratio larger than 20, up to 150 keV.  The grouped spectra 
from the four units were then coadded.  

Spectral analysis was performed with the XSPEC 9.01 package, using for each 
instrument the response matrices released by the SDC.  LECS data have been 
considered only in the range 0.1--4 keV, due to still unsolved calibration 
problems at higher energies, and MECS data in the range 1.8--10.5 keV.  For
each observation, the LECS and MECS spectra have been jointly fit after 
allowing for a constant, systematic rescaling factor of the LECS data, 
to account for uncertainties in the intercalibration of the instruments 
(Parmar 1997), which had a best fit value of 0.64.  We considered both 
simple and broken power-law models; the latter is appropriate for HBL, 
which often show downward curved spectra.  With the former model, the fitted 
$N_{\rm H}$ is 30-40\% higher than the Galactic value ($1.73 \times 10^{20}$ 
cm$^{-2}$; Elvis, Lockman, \& Wilkes 1989), while the broken power-law model 
yields a value equal to the Galactic one for the first two observations and a 
somewhat lower value ($\sim$ 20\%) for the third one.  We then fixed  
$N_{\rm H}$ at the Galactic value and determined the best-fit parameters for 
both the single and broken power-law models (see Table~1): the latter is 
clearly in all cases a better representation of the data (the $\chi^2$ 
improvement is highly significant as estimated from the F-test).  The 
spectral steepening between the LECS and MECS bands is of 
$\Delta\alpha \simeq 0.2-0.3$ at all epochs.  Single power-laws represent 
well the PDS spectra in the 13--200 keV range (see fit parameters in Table~1)
at each epoch.  For 7 and 11 April the PDS slopes are consistent with the 
energy indices derived from the broken power-law LECS+MECS best-fits 
($\alpha_2$ in Table~1), while for 16 April the spectrum in the 13--200 keV 
band is significantly steeper than in the MECS  band.  After rescaling the
PDS data by a factor 0.75 to take into account a calibration mismatch with 
respect to the MECS response, we fit the 16 April spectrum over the whole 
range (LECS+MECS+PDS) with a broken power-law model.  The PDS data show
a systematic deviation from the model (Fig. 1a), indicating a further 
steepening at higher energies.  Figure 1b shows a joint fit only to the 
MECS+PDS data with a broken power-law with break energy of $\sim 20$ keV:
the spectral indices on the lower and higher energy side of the break (see 
figure caption) are similar to $\alpha_2$ and $\alpha_{PDS}$ (Table 1), 
respectively. 

In order to characterize the spectral variability in the full range 
independently of calibration uncertainties or model assumptions, we computed 
the ratio of the count rates observed on 16 April to those observed on 7 
April vs. energy, as shown in Figure 2.  It is clear that the variability is 
systematically larger at higher energies implying an overall hardening of the 
spectrum with increasing intensity, by about $\Delta\alpha \simeq 0.2$
(see Table 1).  On 7 and 11 April the spectral index in the PDS band is 
consistent with $\alpha \simeq 1$, while on 16 April it is less than unity, 
indicating that the peak of the power output falls in the PDS band at the 
first two epochs but is at the extreme end of the PDS band or beyond in the 
highest intensity state. 

\section{Spectral energy distribution}

In Figure 3 the unfolded and unabsorbed X-ray spectra from the BeppoSAX
observations of 7 and 16 April 7 are compared with previously observed X-ray 
spectra in three brightness states and with data from the radio to the TeV 
range (see figure caption for references).  The present X-ray observations 
imply a dramatic hardening of the spectral energy distribution in the medium 
X-ray band and an increase of the (apparent) bolometric luminosity of a 
factor $\ge 20$ with respect to previous epochs.  The really striking feature 
is that the peak of the power output (i.e., where $\alpha = 1$) is found to 
{\it have shifted} in energy by a factor $\ge 100$.  Moreover for the first 
time in any blazar the peak is observed to {\it occur in the hard X-ray 
range}, definitely above $\sim 50-100$ keV (cf. Ulrich, Maraschi, \& Urry 
1997).  Since in the optical the source was nearly normal (Buckley \& McEnery 
1997) the change of the spectral energy distribution seems to be confined to 
energies greater than $\sim$0.1 keV, as also indicated by the apparent pivot 
of the three BeppoSAX spectra.  The overall continuity of the X-ray spectra 
reported here with previous UV and soft X-ray measurements suggests that the 
X-ray emission constitutes the high energy end of the synchrotron component
and thus that its peak frequency increased by more than two orders of 
magnitude and its power by more than one order of magnitude with respect to 
previous observations of Mkn~501.

The TeV emission also brightened by more than a factor of 5 in the first two
weeks of April, with the most intense TeV flare peaking on 16 April (Catanese
et al. 1997), the date of the last BeppoSAX observation.  However, unlike the
X-ray spectrum, the TeV spectrum was steeper than $\alpha = 1$ and did not 
show noticeable temporal variation; from the HEGRA measurements during the 
period March-April and during the more active period 7-13 April, the spectral 
index above 1 TeV remained unchanged, within the large errors ($\Delta \alpha 
\sim 0.3$), with an average value $\alpha \simeq 1.5$ (Aharonian et al. 
1997a,b).  At the same time, Mkn~501 was not detected by EGRET (Catanese et 
al. 1997), indicating that the $\gamma$-ray flare is modest at a few GeV.  
This constrains the peak power output of the very high energy component to
occur between the GeV and TeV ranges.  Note that also during the quiescent 
state the latter peak was poorly constrained so that it is difficult to make 
a strong statement about a possible shift. In the quiescent state the TeV 
flux was a factor $\sim$100 less than during the flare of 16 April. 
 
\section {Model implications}

The spectral energy distribution of HBL can be well explained by the 
synchrotron self-Compton model, in which the dominant source of seed 
photons is the synchrotron emission (Jones, O'Dell, \& Stein 1974; 
Ghisellini, Maraschi, \& Dondi 1996; Mastichiadis \& Kirk 1997).  If the
energy distribution of the emitting electrons, $N(\gamma)$, changes at the 
highest energies, this model explains naturally the correlated flaring at 
X-ray and TeV energies, with the highest energy electrons producing X-rays 
via synchrotron and the TeV radiation via inverse Compton scattering.  
Because of the very high electron energies involved, the scattering cross 
section for energetic photons is reduced by the Klein-Nishina effect and only 
photons below the Klein-Nishina threshold ($h\nu\leq mc^2/\gamma$) are 
effectively upscattered.  This means that (i) the inverse Compton flux does 
not vary more than the synchrotron flux from the same electrons, since the 
available seed photons are limited (Ghisellini \& Maraschi 1996) and (ii) the 
peak of the inverse Compton power shifts less in frequency than the 
synchrotron peak.

The shift of $\sim$2 orders of magnitude in the frequency $\nu_S$ of the 
synchrotron peak of Mkn 501 cannot be ascribed to either a variation of 
Doppler factor, $\delta$, or magnetic field, $B$, alone; enormous variations 
would be demanded, since $\nu_S\propto B\delta$, and these changes would 
affect other parts of the spectrum quite strongly in a way that is not
observed.  Therefore, a real change in power and an increase in maximum
electron energy, $\gamma_{max}$, is implied.  Assuming that the power 
variation is only due to a change in the electron energy, $\gamma_{max}$
must have increased by roughly a factor of $\sim$10--30.  The corresponding 
shift of the inverse Compton peak is expected to be of the same order of 
magnitude, being in the Klein-Nishina regime.  Since the cooling time of 
these very high energy electrons is rather short and the synchrotron peak 
did not move back to the quiescent position during at least 10 days, a 
mechanism of continuous particle injection is required.  This could be
impulsive acceleration of electrons at a shock, which might be triggered by 
fluctuations in the velocity or energy of newly injected plasma.
In the $\gamma$-ray emitting region, the fresh electrons would therefore
scatter, besides the synchrotron photons, also a pre-existing (and more 
stationary) photon population, produced by an older electron distribution. 
The hard X-ray and TeV emission should then vary with similar amplitude, 
while the flux in the infrared-optical band could remain almost stationary.

Three one-zone synchrotron self-Compton models along these lines are shown 
in Figure 3 for the quiescent state and the 7 and 16 April states.  The 
$N(\gamma)$ distribution has been found self-consistently solving the 
continuity equation including continuous injection of relativistic particles, 
radiative losses, electron-positron pair production and taking into account 
the Klein-Nishina cross section (Ghisellini 1989).  The parameters of the fits
are given in the figure caption.  A variation of the maximum energy of the 
emitting electrons, together with an increased luminosity and a flattening of 
the injected particle distribution can describe the observed flaring spectra 
quite well.  We assumed that the seed photons for the inverse Compton 
scattering are the sum of those produced by the injected electrons plus a 
stationary component, comparable to the observed infrared-optical flux, which 
is assumed to be cospatial with the high energy emission.  This component can
be associated with the ``quiescent'' spectrum.  We recall that these one-zone 
models cannot account for the radio emission, thought to be produced in much 
larger regions of the source.

We note that the soft X-ray flux (up to a few keV) did not vary dramatically,
and soft X-ray observations alone would have failed to recognize an unusual 
behavior, except for measuring a spectral inversion, namely a change in slope 
from values larger than 1 to values smaller than 1.  In the EXOSAT data base 
(Sambruna et al. 1994) at least two sources (out of 16) show spectral indices
smaller than 1 in the medium energy X-ray range, but in the brightest and most
frequently observed sources (i.e., with more than 10 spectra, which is the 
case also for Mkn~501) such behavior was never observed.  Thus flares as 
discovered in Mkn~501 may occur in other (similar) blazars though not very 
frequently and/or some blazars may be more often in ultra-hard states.  The 
signature would be a flat slope in the X-rays, with a flux level consistent 
with the extrapolation of the infrared-optical synchrotron spectrum.  These 
sources should be the most variable in hard X-rays and the strongest ones in 
the TeV band, therefore, their investigation with the existing and rapidly 
developing X- and $\gamma$-ray instrumentation is one of the most promising 
projects of high energy astronomy.
 
\acknowledgements
We thank F. Aharonian for a critical reading of the manuscript and for
providing helpful comments and suggestions, and M. Catanese for sending us, 
on behalf of the Whipple team, his manuscript in advance of submission.  
J. Buckley, G. Fossati, K. O'Flaherty, M. Schubnell, T. Weekes are 
acknowledged for their support of this project.  CMU and Elena Pian 
acknowledge support from NASA grants NAG5-2510 and NAG5-2538.



\newpage

\cl{\bf Figure Captions}

\figcaption{{\it a)} Broken power-law fit to the LECS+MECS+PDS data of 16
April 1997.  The top panel shows the spectra from each instrument (see Table
1 for the parameters); the bottom panel shows the ratio between the spectral 
flux distribution and the model; {\it b)} same as in {\it (a)} for the 
MECS+PDS data of 16 April: best fit spectral indices at energies lower and 
higher than the break energy of $19.5^{+4.5}_{-3.5}$ keV are 
$0.58^{+0.03}_{-0.02}$ and $0.83^{+0.08}_{-0.06}$, respectively 
($\chi^2_r=1.24$ for 113 degrees of freedom).} 

\figcaption{Ratio of the LECS+MECS+PDS spectra of 7 and 16 April.
The hardening of the 16 April spectrum is clearly evident.}

\figcaption{Spectral energy distribution of Mkn~501.  BeppoSAX data from 7 
and 16 April 1997 are indicated as labeled.  Nearly simultaneous Whipple 
TeV data (from Catanese et al. 1997) are indicated as filled circles, while
the open circle (13 April 1997) and the TeV spectral fit (15-20 March 1997) 
along with its 1-$\sigma$ confidence range are from the HEGRA experiment 
(Aharonian et al. 1997a).  Non-simultaneous measurements collected from the 
literature are shown as open squares (radio, Gear et al. 1994; millimeter, 
Steppe et al. 1988, Wiren et al. 1992, Lawrence et al. 1991, Bloom \& 
Marscher 1991; far-infrared Impey \& Neugebauer 1988; optical, V\'eron-Cetty 
\& V\'eron 1993, Burbidge \& Hewitt 1987; UV, Pian \& Treves 1993; TeV, Quinn
et al.  1996, Breslin et al. 1997, Bradbury et al. 1997).  X-ray spectral 
fits in the low state are from Sambruna et al. (1994), Worrall \& Wilkes 
(1990), Comastri et al. (1997).  Upper limits at 100 MeV are from Weekes et 
al. (1996), Catanese et al. (1997).  The solid lines indicate fits with a 
one-zone, homogeneous synchrotron self-Compton model. For all models the size
of the emitting region is $R = 5\times 10^{15}$ cm, the beaming factor is 
$\delta = 15$ and the magnetic field is $B\sim 0.8$ Gauss.  For the 
``quiescent state'', the intrinsic luminosity (corrected for beaming) is 
$L^\prime = 4.6\times 10^{40}$ erg s$^{-1}$, and electrons are continuously 
injected with a power-law distribution ($\propto \gamma^{-2}$) between 
$\gamma_{min}=3\times 10^3$ and $\gamma_{max}=6\times 10^5$.  For the fit to 
the 7 April spectrum, $L^\prime = 1.8\times 10^{41}$ erg s$^{-1}$ and the 
injected electron distribution ($\propto \gamma^{-1.5}$) extends from 
$\gamma_{min} ^4$ to $\gamma_{max}=3\times 10^6$.  For the fit to the 16 
April spectrum, $L^\prime = 5.5\times 10^{41}$ erg s$^{-1}$ and the injected 
electron distribution ($\propto \gamma^{-1}$) extends from 
$\gamma_{min}=4\times 10^5$ to $\gamma_{max}=3\times 10^6$.  For the 7 and 
16 April models, the seed photons for the Compton scattering are the sum of 
those produced by the assumed electron distribution plus those corresponding
to the quiescent spectrum.} 

\vfill
\eject
\begin{center}
\begin{tabular}{llll}
\multicolumn{4}{c}{\bf Table 1: Spectral fits$^a$ to LECS+MECS and PDS data}\\
\hline
\hline
Obs. Start-End$^b$ & 7.2295-7.6679 &  11.2458-11.6846 & 16.1506-16.6082 \\
$\alpha^c$ & $0.80 \pm 0.01$ & $0.74 \pm 0.01$ & $0.52 \pm 0.01$ \\
$\chi^2_r$ (N$_{d.o.f.}$) & 3.4 (180) &  1.78 (180) & 2.77 (180) \\
$\alpha_1^d$ (E $<$ E$_{break}$) & $0.63 \pm 0.04$ & $0.64^{+0.02}_{-0.04}$ &
     $0.40^{+0.02}_{-0.04}$ \\      
$\alpha_2^d$ (E $>$ E$_{break}$) & $0.91 \pm 0.02$ & $0.80 \pm 0.02$ &
     $0.59^{+0.02}_{-0.01}$ \\
E$_{break}$ (keV) & $1.76^{+0.24}_{-0.22}$ & $1.85 \pm 0.33$ & 
     $2.14^{+0.30}_{-2.14}$ \\
$\chi^2_r$ (N$_{d.o.f.}$) & 1.29 (178) & 1.00 (178) & 1.33 (178) \\
$\alpha_{PDS}$ & $0.99 \pm 0.13$ & $0.83^{+0.09}_{-0.08}$ &
      $0.84^{+0.03}_{-0.04}$ \\
$\chi^2_r$ (N$_{d.o.f.}$) & 0.92 (12) & 1.60 (12) & 1.09 (12) \\ 
$S_{0.1-2 keV}^e$ & $1.75 \pm 0.01$ & $1.75 \pm 0.01$ & $2.55 \pm 0.01$ \\
$S_{2-10 keV}^e$ & $2.20 \pm 0.01$ & $2.45 \pm 0.01$ & $5.35 \pm 0.01$ \\ 
$S_{13-200 keV}^e$ & $3.75 \pm 0.15$ & $5.15 \pm 0.15$ & $15.8 \pm 0.2$ \\ 
\hline
\multicolumn{4}{l}{$^a$ $F_\nu \propto \nu^{-\alpha}$. Errors are at 90\% 
confidence level for one ($\Delta \chi^2 = 2.71$)}\\
\multicolumn{4}{l}{~~ or three ($\Delta \chi^2 = 6.1$) parameters of 
interest.}\\
\multicolumn{4}{l}{$^b$ Day of April 1997.}\\
\multicolumn{4}{l}{$^c$ Energy index from the single power-law fit to
combined LECS and}\\
\multicolumn{4}{l}{~~ MECS data.}\\
\multicolumn{4}{l}{$^d$ Energy index from the broken power-law fit to
combined LECS and}\\ 
\multicolumn{4}{l}{~~ MECS data.}\\
\multicolumn{4}{l}{$^e$ Unabsorbed intensities, in units of 
$10^{-10}$ erg s$^{-1}$ cm$^{-2}$, derived from}\\ 
\multicolumn{4}{l}{~~ the broken power-law model fit.}\\ 
\end{tabular}
\end{center}
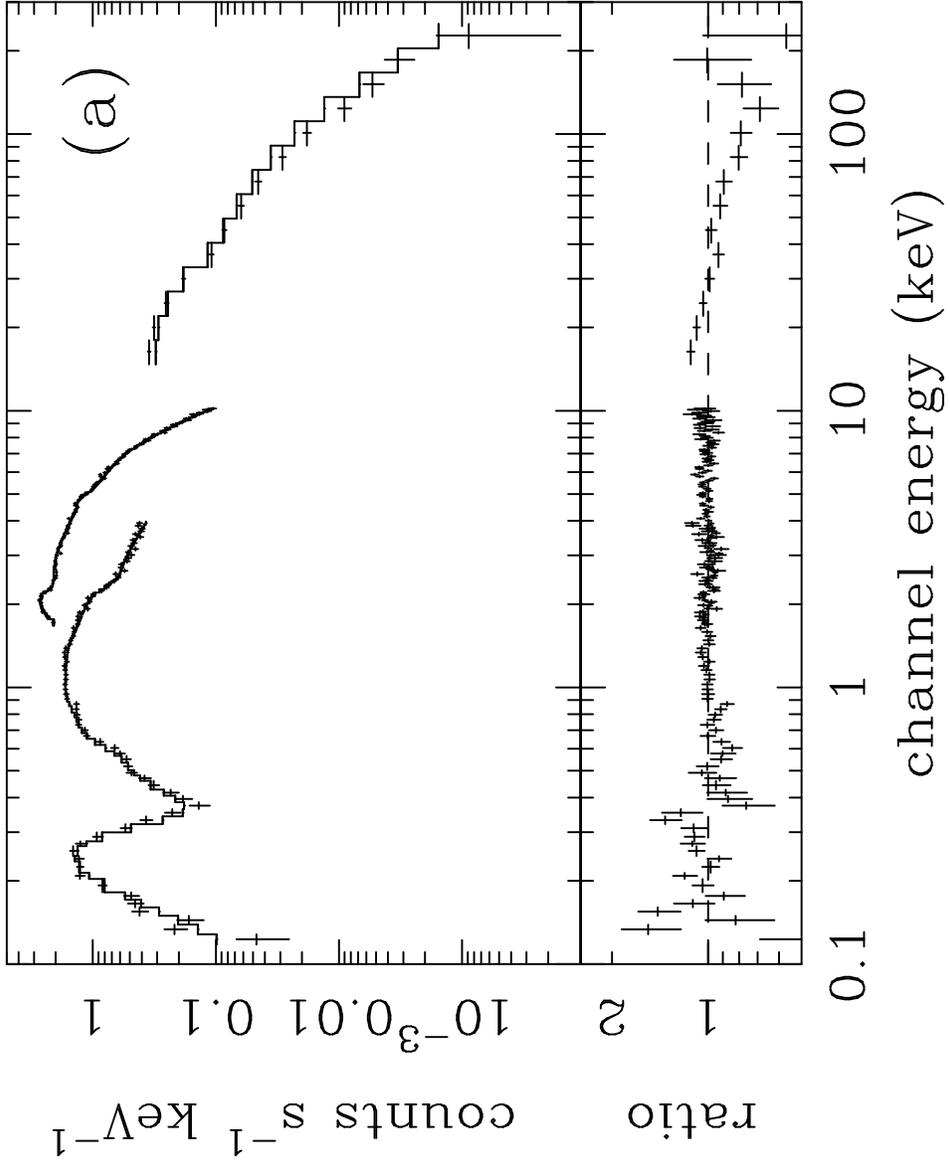
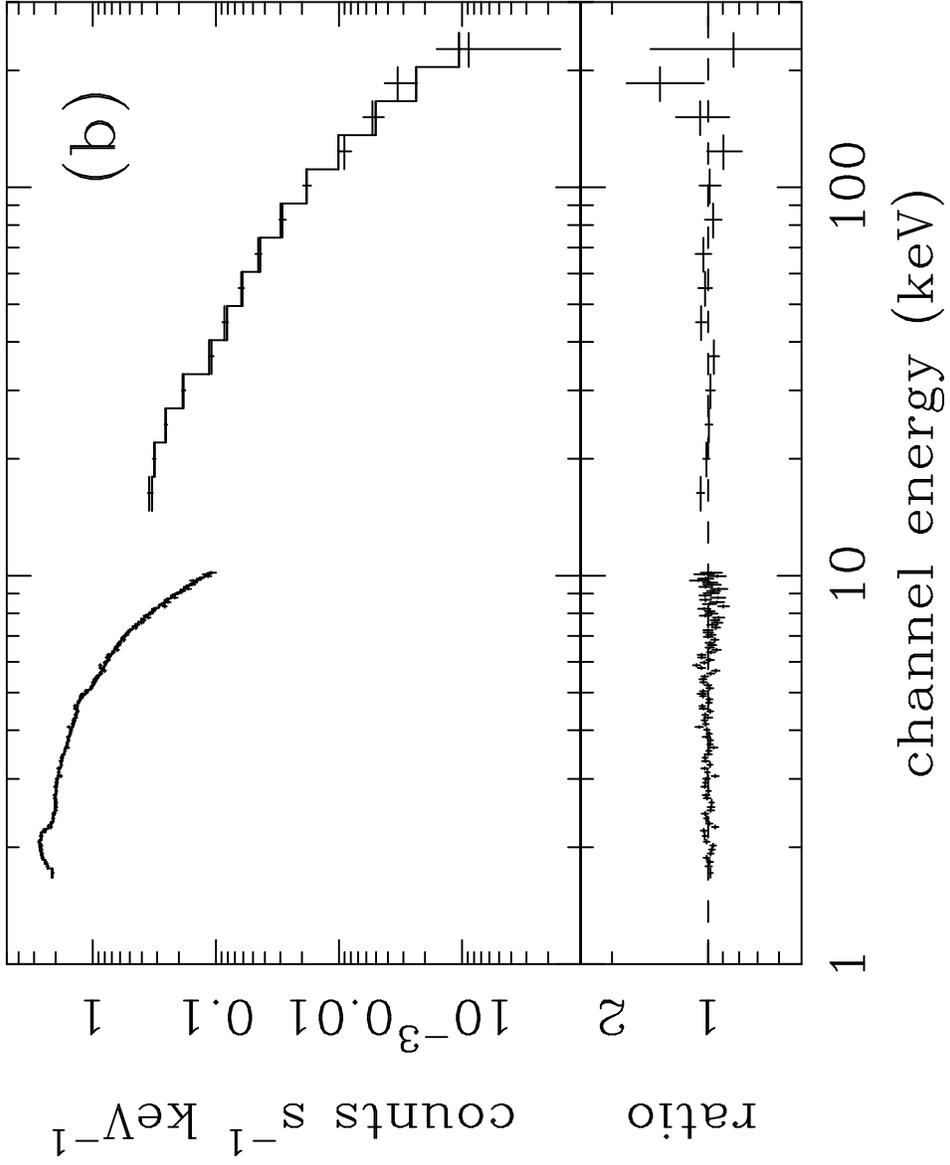
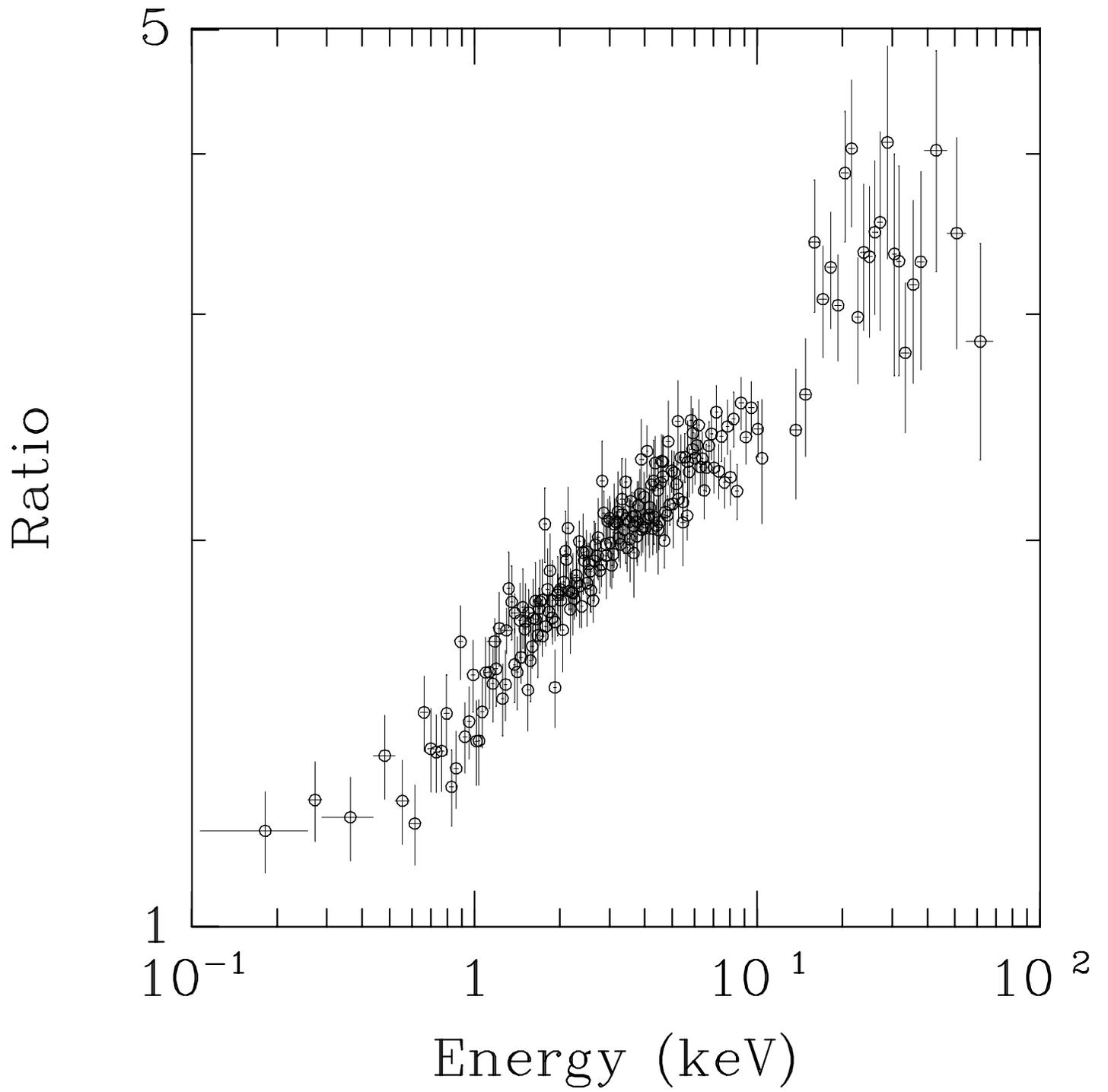
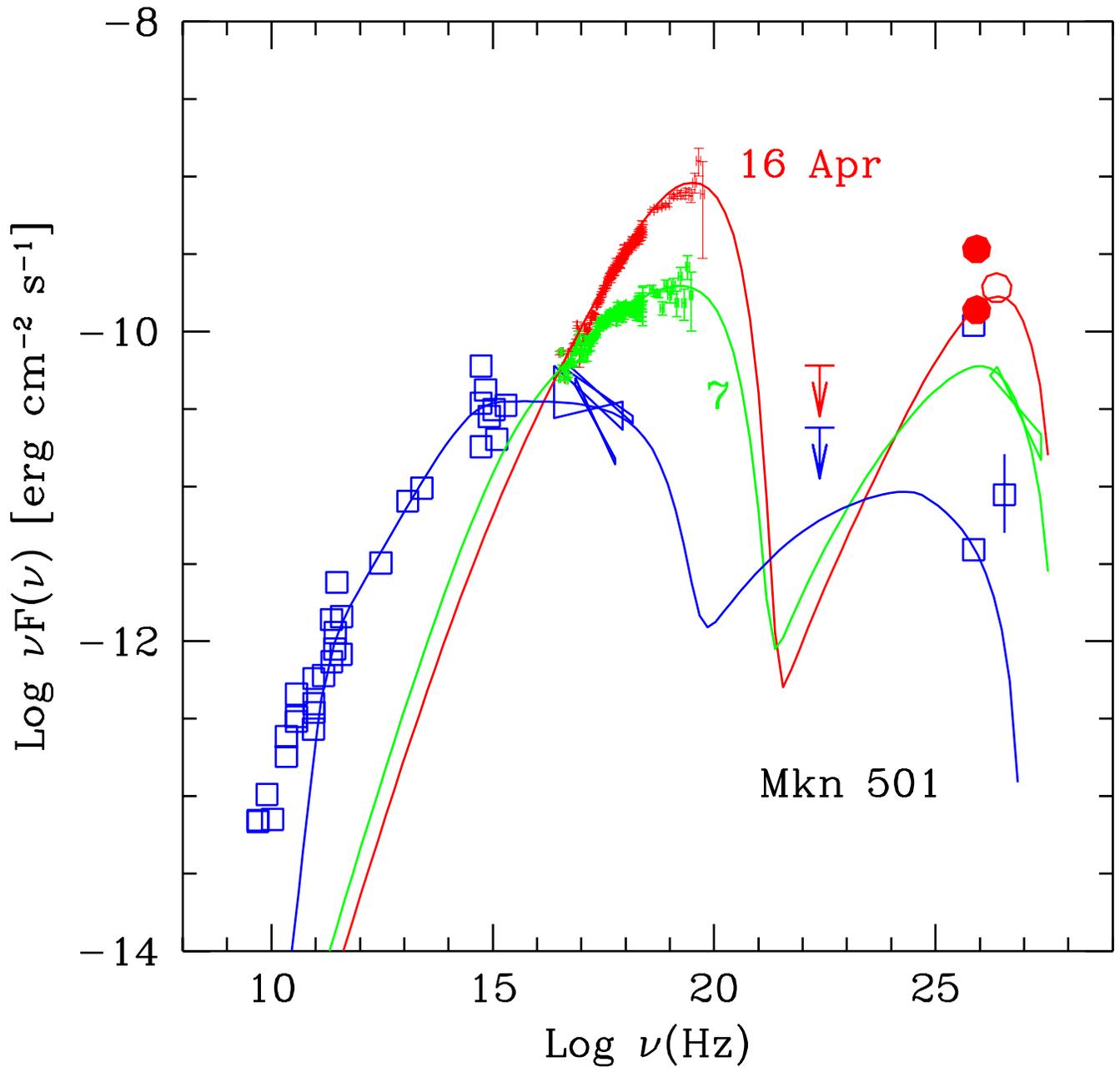
\end{document}